\documentclass[12pt]{amsart}
\usepackage[a4paper, margin=1.2in]{geometry}
\usepackage{amsmath}
\usepackage{amsfonts}
\usepackage{cite}
\usepackage{url}
\usepackage{hyperref}
\usepackage{xurl}

\begin{document}

\title{Metrics Over Merit: The Hidden Costs of Citation Impact in Research}
\author{Vugar Ismailov}

\address{Institute of Mathematics and Mechanics, Baku, Azerbaijan; \newline Center for Mathematics and its Applications, Khazar University, Baku, Azerbaijan}
\email{vugaris@mail.ru}

\begin{abstract}
Once upon a time, scientists' worth was measured by their ideas, proofs, and perhaps how eloquently they debated Hilbert's problems at seminars. But now, citation metrics have come to center stage and handed us new masters: FWCI and CNCI. This paper critically, and with a touch of satire, examines how these seemingly objective metrics are shaping, and often distorting, the scientific landscape. Through examples and analysis, we highlight the consequences of relying too heavily on such indicators in evaluating researchers and scientific contributions.
\end{abstract}

\maketitle

\section{Introduction: The Metrics Era}

In the golden age of mathematical thought, research was judged by its depth, novelty, and influence within the field. Today, it is evaluated by acronyms. Two of the most fashionable ones, FWCI (Field-Weighted Citation Impact) and CNCI (Category Normalized Citation Impact), have made their way from bibliometric backrooms to center stage, shaping decisions on hiring, funding, and rewards for researchers.

These numbers are no longer mere indicators, they have become instruments of judgment. And though they gleam in reports and rankings, their value to real science is often imaginary.

To clarify, let us introduce the protagonists of our story, along with their actual formulas. Note that both of the following metrics are originally defined \textit{at the level of a single publication}:

\textit{FWCI (Field-Weighted Citation Impact)}\cite{FWCI}:
\[
\text{FWCI} = \frac{\text{Actual Citations}}{\text{Expected Citations}}
\]
Here, the \textit{expected citations} are computed based on publications that are similar in terms of subject area, publication year, and document type, according to Scopus data. An FWCI of 1 indicates that the paper has been cited exactly as expected; a value greater than 1 indicates above-average citation impact, while a value below 1 implies below-average performance.

\textit{CNCI (Category Normalized Citation Impact)}\cite{CNCI}:
\[
\text{CNCI} = \frac{\text{Actual Citations}}{\text{Expected Citations}}
\]
This metric uses essentially the same formula as FWCI but relies on \textit{Web of Science categories and data} (Clarivate Analytics) to determine the actual and expected citation count.

Although both FWCI and CNCI are fundamentally defined for \textit{individual articles}, these metrics are used to assess the performance of researchers, departments, institutions, or even countries. In such cases, the overall FWCI or CNCI is typically calculated as the \textit{mean} of the individual values across all considered publications. That is,
\[
\text{Overall FWCI} = \frac{1}{n} \sum_{i=1}^n \text{FWCI}_i, \quad \text{and similarly for CNCI},
\]
where \( n \) is the number of publications. This value is then interpreted in the same way: a value above 1 suggests that, on average, the publications in the group are cited more than expected for their respective fields, years, and document types.

Sounds elegant? Perhaps. But in reality, it often just means rewarding the wrong things. What seems impressive can hide a lot of problems.

This paper was inspired in part by Alberto Saracco's paper \cite{Saracco}, which humorously criticizes the growing obsession with citation counts in academic life. Saracco draws attention to how researchers, under pressure from evaluation systems, may resort to self-citations, group citations, and other tricks to boost their numbers. While his focus is on the raw number of citations, our paper highlights a newer and more subtle problem: the rise of so-called ``normalized" metrics like FWCI and CNCI. These numbers claim to measure the quality of research by comparing it to expected citation levels, but we argue that they introduce even more confusion and unfairness. Our aim is to demonstrate how these metrics influence academic careers, often in ways that have little to do with real scientific merit.

\section{The Illusion of Citation Impact}

As noted above, when measuring a researcher's overall performance, FWCI and CNCI are calculated as the average of each paper's individual score, not the sum. This means that papers with low impact bring down the overall average and can negatively affect the researcher's rating. Unlike total citation counts, which only grow, these metrics can penalize researchers for publishing work with less impact.

Consider a simple scenario. Suppose a researcher has authored 10 papers: three with a FWCI of 2, and seven with a FWCI of 0.1. The overall FWCI? Around 0.67. Now imagine a second researcher who coauthored the first three papers with the first and wrote no other papers. His overall FWCI? A glorious 2.

Who's the better scientist?

According to the metric, it's the second one. The first, despite being more active and possibly contributing more broadly to the field, is punished for publishing more.

Moral of the story? Publish less. Ideally, three high-impact papers, then disappear from the academic stage. Take a sabbatical, write philosophy, teach undergraduates, or go tend a vineyard in silence. Just don't risk lowering your score.

\section{A Case Study: the SASHE Evaluation Criteria}

The State Agency for Science and Higher Education (SASHE) in Azerbaijan has introduced a scoring system to evaluate researchers using a standardized, quantitative framework based on their publications over the past six years. Each researcher receives a total score composed of two major components: a Scopus Score and a Web of Science Score. These are calculated using three main criteria:
\begin{itemize}
\item The researcher's overall FWCI and CNCI;
\item The number of papers ranked among the top 10\% in Scopus and Web of Science, based on their FWCI and CNCI respectively;
\item The number of publications in Q1 or Q2 journals, as designated by the respective databases.
\end{itemize}

Among these, the citation impact metric (FWCI or CNCI) typically carries the greatest weight, followed by the top 10\% indicator, and then the Q1/Q2 journal count.

The intent is clear: to reward researchers whose work is frequently cited and published in journals with high impact factors. However, the system's reliance on citation metrics introduces unintended consequences. Because the evaluation system heavily depends on a researcher's FWCI and CNCI, which are calculated as averages over their publications, adding even a single new paper, particularly if it is uncited or lowly cited and published in a journal outside the Q1/Q2 categories, can reduce the overall citation impact component, thereby lowering the total SASHE score.

This paradox is not unique. Even a publication in a top-tier journal such as \textit{Annals of Mathematics} or \textit{Acta Mathematica}, undisputed pillars of scholarly excellence, may actually reduce a researcher's total score if it has not yet accumulated citations. This counterintuitive outcome occurs because the total score is dominated by the citation impact metric (FWCI or CNCI), while the Q1/Q2 classification contributes comparatively little. Although Annals of Mathematics and similar journals are ranked as Q1, their contribution to the total score is relatively minor. Meanwhile, if the newly published article is still uncited, it reduces the researcher's FWCI or CNCI, the most heavily weighted component. Furthermore, since the paper has not entered the top 10\% category, it contributes nothing to that component either. As a result, the total SASHE score may decline, despite the publication being in one of the most prestigious venues in mathematics.

\section{How to Win at Metrics (by Doing Less)}

Once the new metrics overlords, FWCI and CNCI, begin to shape careers, researchers adjust their behavior to match the rules of the game.

Imagine a researcher with a few well-cited papers and strong FWCI and CNCI. Then comes another paper, published in a top journal, reflecting deep theoretical work that has not yet found its audience. Citations trickle in slowly or not at all. The metrics drop. The numbers whisper: you should have stopped while you were ahead. Worse, the system draws no distinction between ``not cited yet" and ``not worth citing." Time and context are ignored.

The consequences are predictable. Researchers quickly learn to:
\begin{itemize}
\item[(a)] Avoid topics that take time to gain recognition;
\item[(b)] Focus on subfields where citations accumulate rapidly;
\item[(c)] Delay or abandon promising but citation-risky projects;
\item[(d)] Co-author frequently within strategic networks to benefit from mutual referencing practices;
\item[(e)] Publish in journals known for rapid review and high citation rates rather than in high-quality venues where impact may come slowly;
\item[(f)] Consistently cite their own previous publications to inflate cumulative citation metrics;
\item[(g)] Intentionally include minor, correctable errors to justify publishing corrigenda that cite the original work.
\end{itemize}

These are not fringe cases; they are logical responses to the incentives. Under this regime, the safest strategy is to write a few strategic papers, ensure they are cited quickly, and then retreat into academic hibernation.

Metrics are meant to reward excellence. Yet when they punish productivity and genuine effort, caution takes over, because meaningful deep work can indeed become a risk to your score.

\section*{Acknowledgments}

I would like to thank the metrics overlords, FWCI, CNCI, and their rapidly multiplying family, for providing the inspiration to write this paper. In many ways, it is a direct response to the academic behaviors these metrics now shape.

\section*{A Final, Serious Note}

While parts of this paper adopt a critical or ironic tone, the problem at hand is not a joke. Citation-based metrics like FWCI and CNCI are now regularly used to evaluate researchers, departments, and institutions, despite being poorly understood by most of the people affected by them.

These numbers create an illusion of fairness and objectivity, but in practice they penalize certain fields and types of research. They reward research that attracts attention rather than research that has real depth, favor popular topics instead of new ideas, and favor what fits current demands over true scientific contribution. Worse still, they push researchers to adapt their behavior, not to improve the quality of their work, but to satisfy opaque scoring formulas.

If we want a healthy scientific ecosystem, we must stop treating FWCI, CNCI, and similar indicators as if they were measures of intellectual merit. They are not. At best, they are administrative conveniences. At worst, they are obstacles to real progress.

\textit{Metrics should serve science, not the other way around.}

\

\end{document}